\documentclass[aps,prl,twocolumn,floats,balancelastpage,showpacs,showkeys,preprintnumbers,floatfix,nofootinbib,superscriptaddress]{revtex4}
\usepackage{graphicx,amsmath,amssymb,amsfonts, soul}
\usepackage{hyperref}

\begin{document}
\title{Search for dark matter annihilations towards the inner Galactic halo \\ from 10 years of observations with H.E.S.S.}
\author{H.E.S.S. Collaboration}
\email[Correspondence should be sent to ]{contact.hess@hess-experiment.eu}
\noaffiliation

\author{H.~Abdallah} 
\affiliation{Centre for Space Research, North-West University, Potchefstroom 2520, South Africa}

\author{A.~Abramowski}
\affiliation{Universit{\"a}t Hamburg, Institut f{\"u}r Experimentalphysik, Luruper Chaussee 149, D 22761 Hamburg, Germany}

\author{F.~Aharonian}
\affiliation{Max-Planck-Institut f{\"u}r Kernphysik, P.O. Box 103980, D 69029 Heidelberg, Germany}
\affiliation{Dublin Institute for Advanced Studies, 31 Fitzwilliam Place, Dublin 2, Ireland}
\affiliation{National Academy of Sciences of the Republic of Armenia, Marshall Baghramian Avenue, 24, 0019 Yerevan, Republic of Armenia}

\author{F.~Ait Benkhali}
\affiliation{Max-Planck-Institut f{\"u}r Kernphysik, P.O. Box 103980, D 69029 Heidelberg, Germany}

\author{A.G.~Akhperjanian}

\affiliation{National Academy of Sciences of the Republic of Armenia, Marshall Baghramian Avenue, 24, 0019 Yerevan, Republic of Armenia}
\affiliation{Yerevan Physics Institute, 2 Alikhanian Brothers St., 375036 Yerevan, Armenia}

\author{E.~Ang{\"u}ner}
\affiliation{Institut f{\"u}r Physik, Humboldt-Universit{\"a}t zu Berlin, Newtonstr. 15, D 12489 Berlin, Germany}

\author{M.~Arrieta}
\affiliation{LUTH, Observatoire de Paris, PSL Research University, CNRS, Universit\'e Paris Diderot, 5 Place Jules Janssen, 92190 Meudon, France}

\author{P.~Aubert}
\affiliation{Laboratoire d'Annecy-le-Vieux de Physique des Particules, Universit{\'e} de Savoie, CNRS/IN2P3, F-74941 Annecy-le-Vieux, France}

\author{M.~Backes}
\affiliation{University of Namibia, Department of Physics, Private Bag 13301, Windhoek, Namibia}

% ...

\begin{abstract}
The inner region of the Milky Way halo harbors a large amount of dark matter (DM). Given its proximity, 
it is one of the most promising targets to look for DM. 
We report on a search for the annihilations of DM particles using $\gamma$-ray
observations towards the inner 300 parsecs of the Milky Way, with the
H.E.S.S. array of ground-based Cherenkov telescopes.
The  analysis is based on a 2D maximum likelihood method
using Galactic center (GC) data  accumulated by H.E.S.S. over the last 10 years (2004-2014), and does not show any significant $\gamma$-ray signal above background. Assuming Einasto and Navarro-Frenk-White DM density profiles at the GC,  we derive upper limits on the annihilation cross section $\langle \sigma v\rangle$. These constraints are the strongest obtained so far in the TeV DM mass range and improve upon previous limits by a factor 5. For the Einasto profile, the constraints reach $\langle \sigma v\rangle$ values  of $\rm 6\times10^{-26} cm^3s^{-1}$ in the $W^+W^-$ channel for a DM particle mass of 1.5 TeV, and $\rm 2\times10^{-26} cm^3s^{-1}$ 
in the $\tau^+\tau^-$ channel for 1 TeV mass.
For the first time, ground-based $\gamma$-ray observations have reached sufficient sensitivity to probe $\langle \sigma v\rangle$ values expected from the thermal relic density for TeV DM particles.
 \end{abstract}


\maketitle 

\section{Introduction}
\label{sec:introduction}  
About 85\% of the mass content of the universe is composed of cold non-baryonic dark matter (DM)~\cite{Adam:2015rua}. There are many well-motivated elementary particle candidates arising in extensions of the standard model of particle physics. One of the most compelling classes of models assumes DM to consist of weakly interacting massive particles (WIMPs)~\cite{Jungman:1995df,Bergstrom:2000pn,Bertone:2004pz}: stable particles with masses and coupling strengths at the electroweak scale, and produced in a standard thermal history of the Universe have the relic density that corresponds to that of observed DM.  WIMPs would today self-annihilate in high DM density regions producing standard model particles, including a potential continuum emission of very-high-energy (VHE, E$_{\gamma}$ $\gtrsim$ 100 GeV) $\gamma$-rays in the final state that can be detected by the H.E.S.S. (High Energy Stereoscopic System) array of ground-based Cherenkov telescopes.

Observational strategies to search for DM annihilation signals focus on regions in the sky with both 
expected high DM density and reduced astrophysical $\gamma$-ray signals. 
VHE $\gamma$-ray observations of the Galactic center (GC) region are amongst the most  promising avenues to look for DM annihilation signals due to the GC proximity and its expected large DM content. 
DM annihilation signals from the GC are expected to be stronger than those from dwarf galaxies by several orders of magnitudes. However, contrary to the case of dwarf galaxies, observations of the GC region face strong astrophysical backgrounds.
Indeed, the inner 300 parsecs of the GC harbors VHE emitters, namely  the central $\gamma$-ray source HESS J1745-290~\cite{Aharonian:2004wa,Aharonian:2009zk}, the supernova/pulsar wind nebula G0.9+0.1~\cite{Aharonian:2005br}, the supernova remnant HESS J1745-303~\cite{Aharonian:2008gw}, and a diffuse emission extending along the Galactic plane~\cite{Aharonian:2006au}.  For DM particles in the TeV mass range, the strongest constraints on the velocity-weighted annihilation cross section $\langle \sigma v\rangle$  to date lie at $\rm{3 \times 10^{-25}\ cm^{3}s^{-1}}$ from 112 hours of observation towards the GC region by H.E.S.S. and a parametrisation of the $\gamma$-ray annihilation spectrum via $q\bar{q}$ pairs~\cite{Abramowski:2011hc}.

The differential $\gamma$-ray flux from the self annihilation of DM particles of mass $m_{\rm DM}$ in a solid angle $\rm \Delta\Omega$ is given by
\begin{equation}
\begin{aligned}
\label{eq:promptflux}
\frac{{\rm d} \Phi_{\gamma}}{{\rm d} E_{\gamma}} (E_{\gamma}, \Delta\Omega) = \frac{\langle \sigma v \rangle}{8\pi \ m_{\rm DM}^2} 
 \frac{{\rm d} N_\gamma}{{\rm d} E_{\gamma}}(E_{\gamma})\times J(\Delta\Omega) \ ,  \\
{\rm with} \, \, \, \,  J(\Delta\Omega)=\int_{\Delta\Omega}\int_{\rm l.o.s.}{\rm d} s \ {\rm d} \Omega \ \rho^2(r(s,\theta))\ . 
 \end{aligned} 
\end{equation}
The particle physics properties are encapsulated in the first term where $\langle \sigma v \rangle$ is the thermally-averaged velocity-weighted annihilation cross section, and ${\rm d} N_{\gamma}/{\rm d} E_{\gamma}(E_{\gamma}) = \sum_f B_{\rm f} \ {\rm d} N^f_{\gamma}/{\rm d} E_{\gamma}(E_{\gamma})$ is the total differential $\gamma$-ray yield per annihilation which corresponds to the sum of the differential $\gamma$-ray yields over the final states $f$ with branching ratio $B_{\rm f}$.\footnote{Only prompt emission of $\gamma$-rays is considered here. Secondary emission is expected from inverse Compton scattering of energetic electrons produced in the DM annihilation on ambient radiation fields. This is particularly relevant in the lepton channels as shown in Ref.~\cite{Lefranc:2015pza}.} The function $J(\Delta\Omega)$, referred to hereafter as the {\it J-factor}, integrates the square of the DM density $\rho$  along the line of sight ($l.o.s.$) in a solid angle $\Delta\Omega$. The coordinate $r$ reads $r = (r^2_{\odot}+s^2-2 r_{\odot} s\ {\rm cos}\ \theta)^{1/2}$, where $s$ is the distance along the line of sight, $\theta$ is the angle between the direction of observation and the
Galactic center, and $r_{\odot}$ is the distance of the observer to the GC assumed to be 8.5 kpc. 
In this analysis, the DM density distribution is parametrized with cuspy profiles. Cored profiles are not considered 
here. They require specific data taking and analysis methods as shown in Ref.~\cite{HESS:2015cda}.
Cuspy profiles are commonly described by Einasto~\cite{Springel:2008by} or Navarro-Frenk-White (NFW)~\cite{Navarro:1996gj} parametrizations, given by
\begin{equation}
\label{eq:profiles}
\begin{split}
\rho_{\rm E}(r) = \rho_{\rm s}  \exp \left[-\frac{2}{\alpha_{\rm s}}\left(\Big(\frac{r}{r_{\rm s}}\Big)^{\alpha_{\rm s} }-1\right)\right]\\
{\rm and} \, \, \, \, \rho_{\rm NFW}(r) = \rho_{\rm s}\left(\frac{r}{r_{\rm s}}\Big(1+\frac{r}{r_{\rm s}}\Big)^2\right)^{-1}  \ , 
\end{split}
\end{equation}
respectively. The Einasto and NFW profile parameters, ($\rho_{\rm s}, \alpha_{\rm s}, r_{\rm s}$) and ($\rho_{\rm s}, r_{\rm s}$), are extracted from Ref.~\cite{Abramowski:2011hc} assuming a local DM density of $\rho_{\odot} = 0.39\ \rm GeV\ cm^{-3}$. 
The J-factors computed in a circular region of 1$^{\circ}$ radius excluding a $\pm$0.3$^{\circ}$ band in Galactic latitudes  to avoid the above-mentioned standard astrophysical emissions, give  $J_{\rm E} =   4.92 \times 10^{21} \ {\rm GeV^2 cm^{-5}}$ and  $J_{\rm NFW} = 2.67 \times 10^{21} \ {\rm GeV^2 cm^{-5}}$ for the Einasto and NFW profiles, respectively. An alternative parametrization of the Einasto profile~\cite{Cirelli:2010xx} leads to $J_{\rm E_2} =   1.51 \times 10^{21} \ {\rm GeV^2 cm^{-5}}$.
The J-factor values for the regions of interest (RoI) considered here  are reported in Table~\ref{tab:jfactors}
 in Supplemental Material~\cite{supplement}.

We reexamine the GC region to search for a DM annihilation signal in the inner Galactic 
halo~\cite{Abramowski:2011hc} using the full statistics from 10 years of GC observations with the initial four telescopes  of the H.E.S.S. instrument~\cite{Aharonian:2006pe}. We perform a new search with an improved data analysis procedure~\cite{2009APh32231D} and a 2-dimensional (2D) likelihood-based method 
using both spectral and spatial characteristics of the DM annihilation signal with respect to background.

\section{Data analysis}
\label{sec:analysis}
The present data analysis makes use of 254 hours (live time) of GC observations during the years 2004-2014 by H.E.S.S. Pointing positions are chosen with radial offsets 
from 0.7$^{\circ}$ to 1.1$^{\circ}$ from the GC. Standard quality selection criteria are applied to the data to select  $\gamma$-ray events~\cite{Aharonian:2006pe}, additionally requiring observational zenith angles
lower than 50$^{\circ}$ to minimize systematic uncertainties in the event reconstruction. The mean zenith angle of the selected observations is 19$^{\circ}$.

The DM signal is analyzed in RoIs defined as annuli of 0.1$^{\circ}$ width each and centered at the GC,
with inner radii from 0.3$^{\circ}$  to 0.9$^{\circ}$ in radial distance from the GC,  
hereafter referred to as the ON regions. 
In order to minimize contamination from the above-mentioned astrophysical emission, 
a band of $\pm$0.3$^{\circ}$ in Galactic latitude is excluded along the Galactic plane.\footnote{Interestingly, 
this enables us to derive constraints that do not strongly depend on the central DM density distribution
which is poorly known in the innermost few tens of parsecs of the GC.} 
The background events are selected in  OFF regions defined  for each observation as annuli symmetric to the ON regions with respect to the pointing position (see Fig.~\ref{fig:background_method}  in Supplemental Material~\cite{supplement}).  OFF regions are expected to contain signal events as well, which decreases any potential excess in the ON regions. 
OFF regions are always taken sufficiently far from the ON regions to obtain a significant contrast in the DM annihilation signal between the ON and OFF regions.\footnote{This analysis method is unable to probe cored profiles (such as isothermal or Burkert profiles). A dedicated observation strategy is required as shown in Ref.~\cite{HESS:2015cda}.}  We considered here the above-mentioned DM profiles for which the OFF regions contain always fewer DM events than the ON regions. A Galactic diffuse emission has been detected by the Fermi satellite~\cite{FermiLAT:2012aa,Ackermann:2014usa}
and H.E.S.S.~\citep{HESS2014sla}. 
Any potential $\gamma$-ray contribution from the Galactic diffuse emission is considered as part of the signal, 
which makes the analysis conservative as long as no signal is detected.
 
We perform a 2D binned Poisson maximum likelihood analysis which 
takes full advantage of the spatial and spectral characteristics of the DM signal with respect to the background. 
We use 70 logarithmically-spaced energy bins from 160 GeV to 70 TeV, and seven spatial bins corresponding to
RoIs defined as the above-mentioned annuli of 0.1$^{\circ}$ width. 
For given DM mass $m_{\rm DM}$ and annihilation channel, the joint likelihood is obtained by the product of the individual Poisson likelihoods over the spatial bins $i$ and the energy bins $j$. It reads:
\begin{equation}
\label{eq:likelihood}
\begin{split}
\mathcal{L}  (m_{\rm DM}, \langle  \sigma v \rangle) = \prod_{\rm i,j} \mathcal{L}_{\rm ij}\ ,   \qquad   \\
\rm{with} \,\,\,\,  \mathcal{L}_{\rm ij}({\mathbf N_{\rm S}}, {\mathbf N_{\rm B}} | {\mathbf N_{\rm ON}}, {\mathbf N_{\rm OFF}}, {\boldsymbol \alpha}) = \\
\frac{\left(N_{\rm S, ij}+ N_{\rm B, ij}\right)^{N_{{\rm ON}, {\rm ij}}}}{N_{{\rm ON}, {\rm ij}}!}e^{-(N_{\rm S,ij}+ N_{\rm B, ij})} \ .
\end{split}
\end{equation}
$N_{\rm S, ij} + N_{\rm B,ij}$ is the expected total number of events in the spatial bin $i$ and spectral bin $j$ of the ON regions. 
The expected number of signal events, $N_{\rm S, ij}$, is obtained by folding the theoretical number of DM events by the instrument response function of H.E.S.S. for this dataset.  $N_{\rm B, ij}$ is the number of background events expected in the spatial bin $i$ and spectral bin $j$.
$N_{\rm ON, ij}$ and  $N_{\rm OFF, ij}$ are the number of observed events in the ON and OFF regions, respectively. $N_{\rm B, ij}$ is extracted from the OFF regions and given by $N_{\rm B, ij} = \alpha_i N_{\rm OFF, ij}$. The parameter $\alpha_i=\Delta\Omega_i/\Delta\Omega_{\rm OFF}$ refers to the ratio between the angular size of the ON region $i$ and the OFF region. In our case, this ratio is equal to one since each OFF region is taken symmetrically to the ON region from the pointing position (including corrections for the camera acceptance). Consequently they have the same angular size and exposure.
$\bf{N_{\rm S}, N_{\rm B}, N_{\rm ON}, N_{\rm OFF}}$ and $\boldsymbol \alpha$ are the vectors corresponding to the quantities previously defined. 
Constraints on $\langle{ \sigma v}\rangle$ are obtained from the likelihood ratio test statistic given by
$ {\rm TS}=-2 \ln(\mathcal{L}(m_{\rm DM},\langle { \sigma v} \rangle)/\mathcal{L}_{\rm max}(m_{\rm DM},\langle { \sigma v} \rangle))$, 
which, in the high statistics limit, follows a $\chi^2$ distribution with one degree of freedom~\citep{Rolke:2004mj}. Values of $\langle { \sigma v}\rangle$ for which TS is higher than 2.71 are excluded at 95\% confidence level (C.L.).

\section{Results}
\label{sec:results}
\begin{figure*}[ht!]
\includegraphics[width=0.45\textwidth]{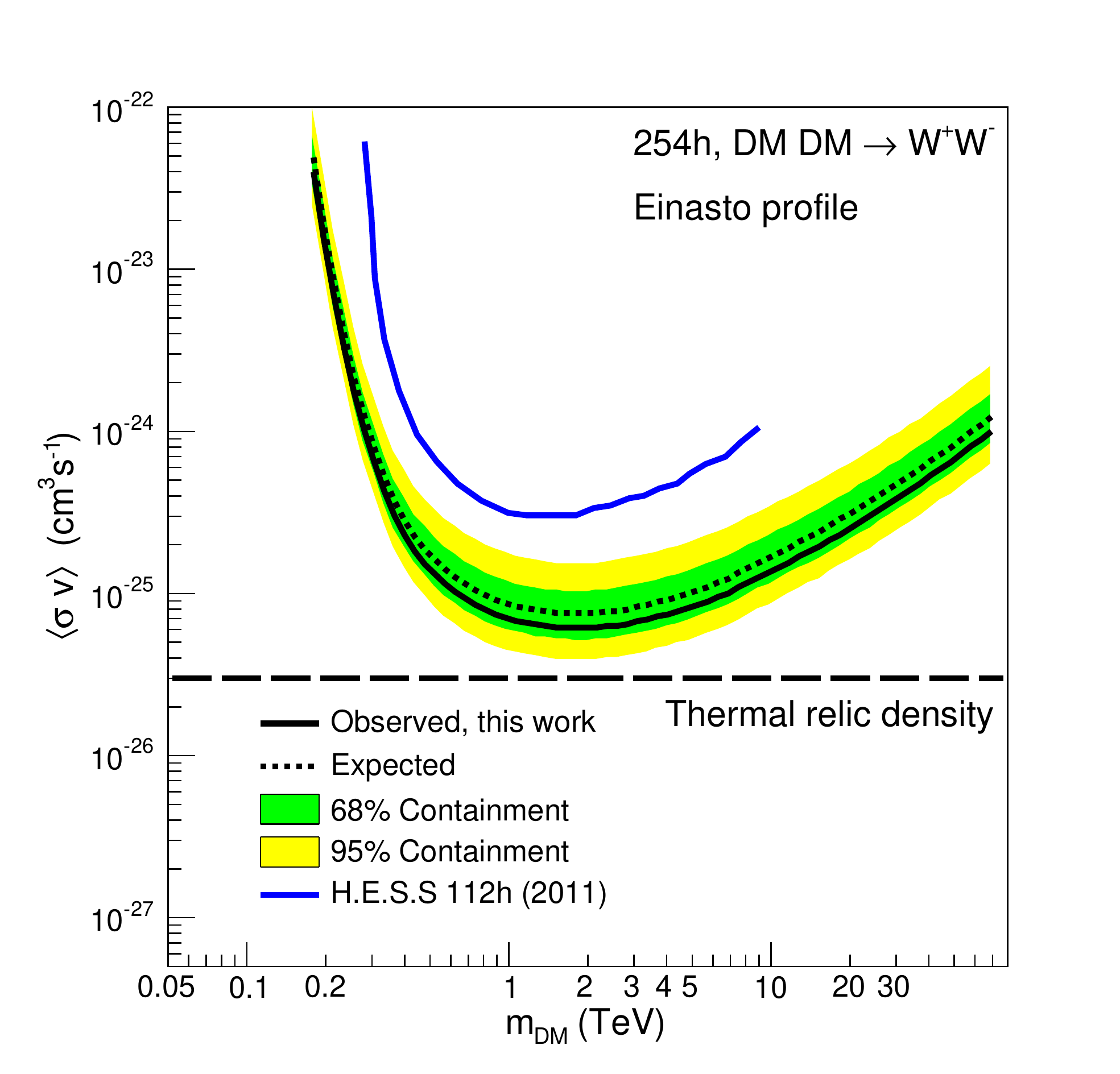}
\includegraphics[width=0.45\textwidth]{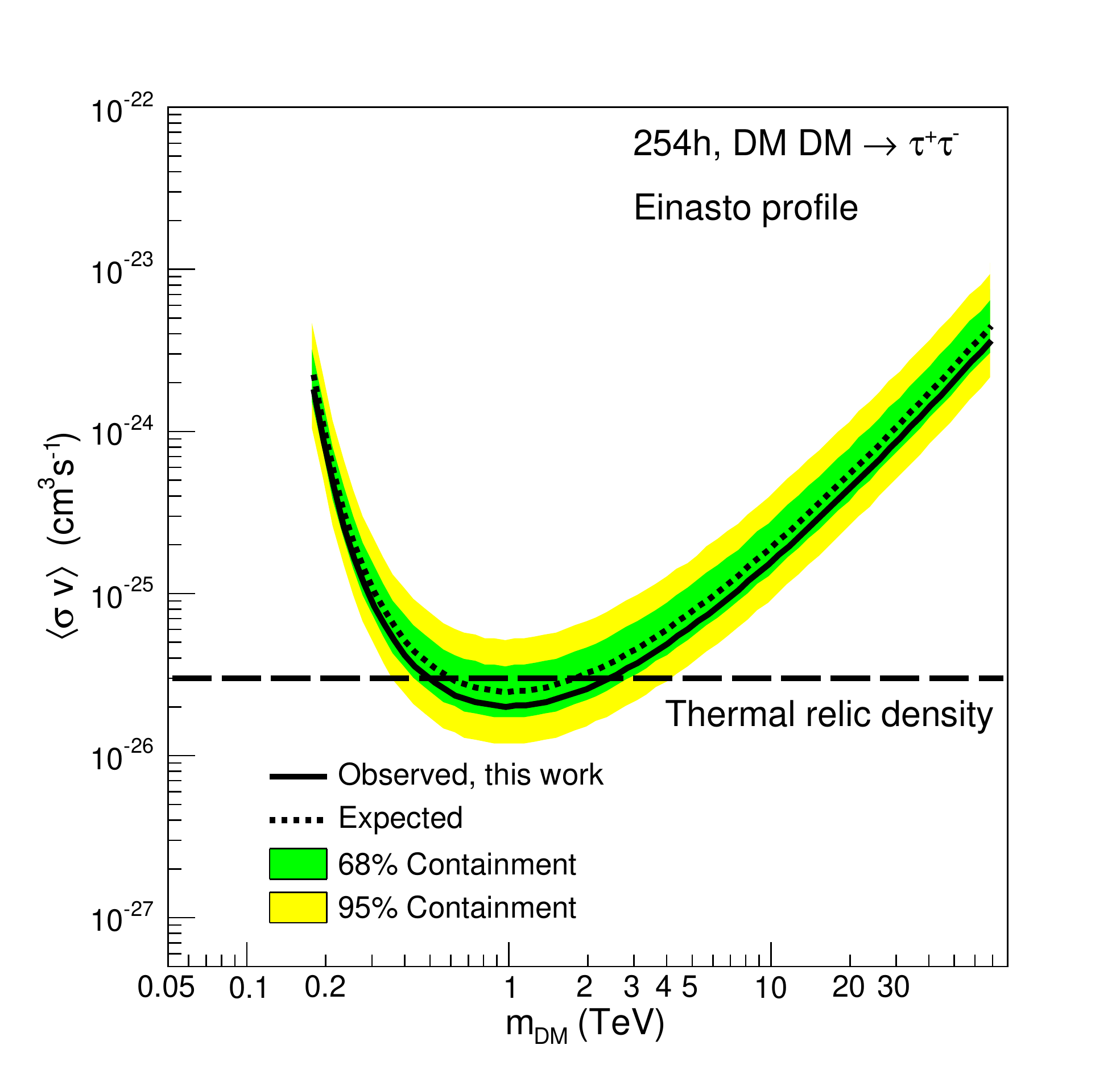}
\caption{Constraints on the velocity-weighted annihilation cross section $\langle \sigma v \rangle$ 
for the  W$^+$W$^-$ (left panel) and  $\tau^+\tau^-$ (right panel) channels derived from observations taken over 10 years of the inner 300 pc of the GC region with H.E.S.S. The constraints for the $b\bar{b}$, $t\bar{t}$ and $\mu^+\mu^-$ channels are given in Fig.~\ref{fig:results_channels} in Supplemental Material~\cite{supplement}.
The constraints are expressed as 95\% C. L. upper limits
as a function of the DM mass m$_{\rm DM}$.  The observed limit is shown as black solid line. The expectations are obtained from 1000 Poisson realizations of the background measured in blank-field observations at high Galactic
latitudes. The mean expected limit (black dotted line) together with the 68\% (green band) and 95\% (yellow band) C. L. containment bands are shown. The blue solid line corresponds to the limits derived in a previous analysis of 4 years (112 h of live time) of GC observations by H.E.S.S.~\cite{Abramowski:2011hc}. 
The horizontal black long-dashed line corresponds to the thermal relic velocity-weighted annihilation cross section (natural scale).  
}
\label{fig:results_errorJ}
\end{figure*}
\begin{figure*}[ht!]
\centering
\includegraphics[width=0.45\textwidth]{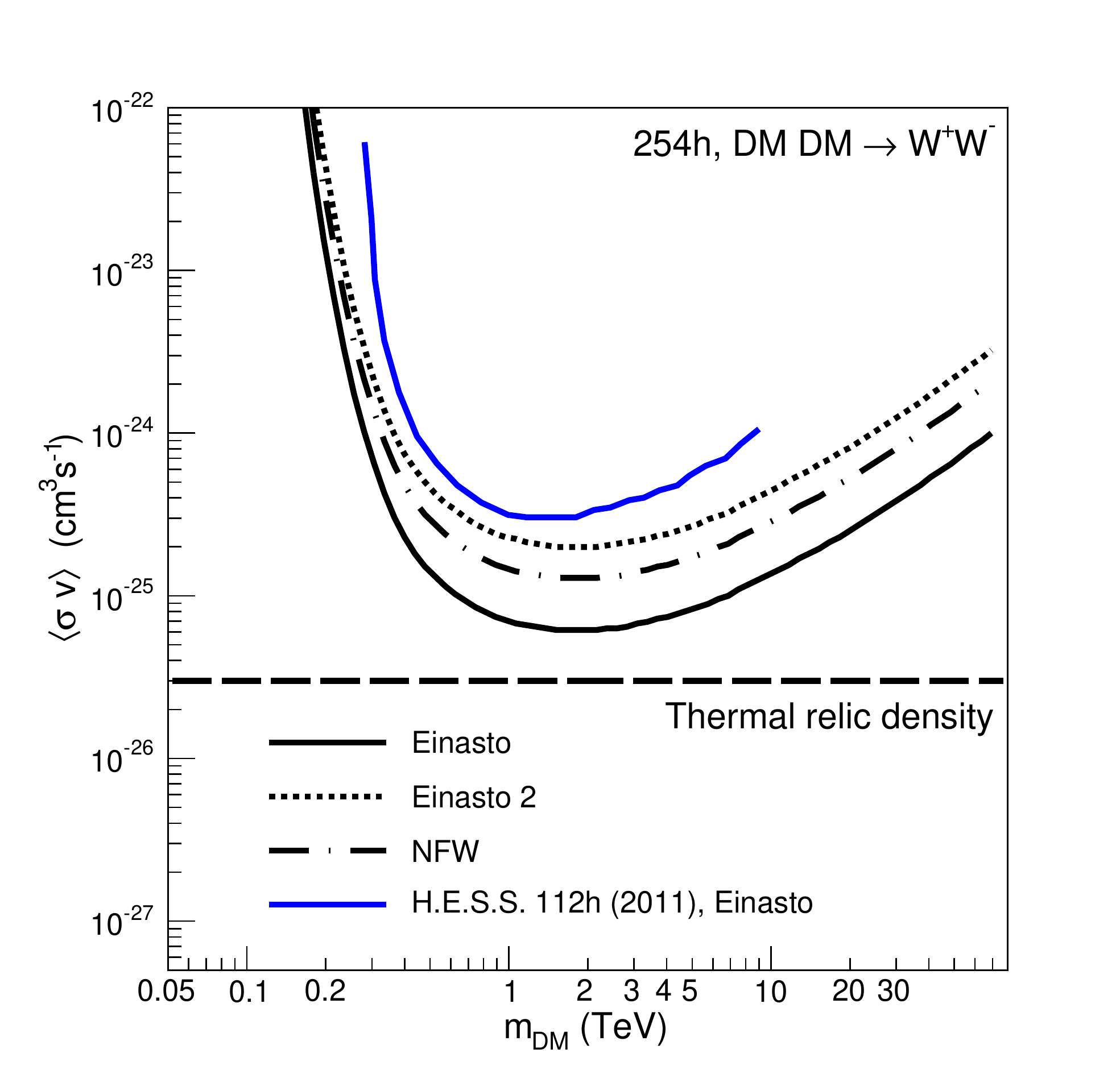}
\includegraphics[width=0.45\textwidth]{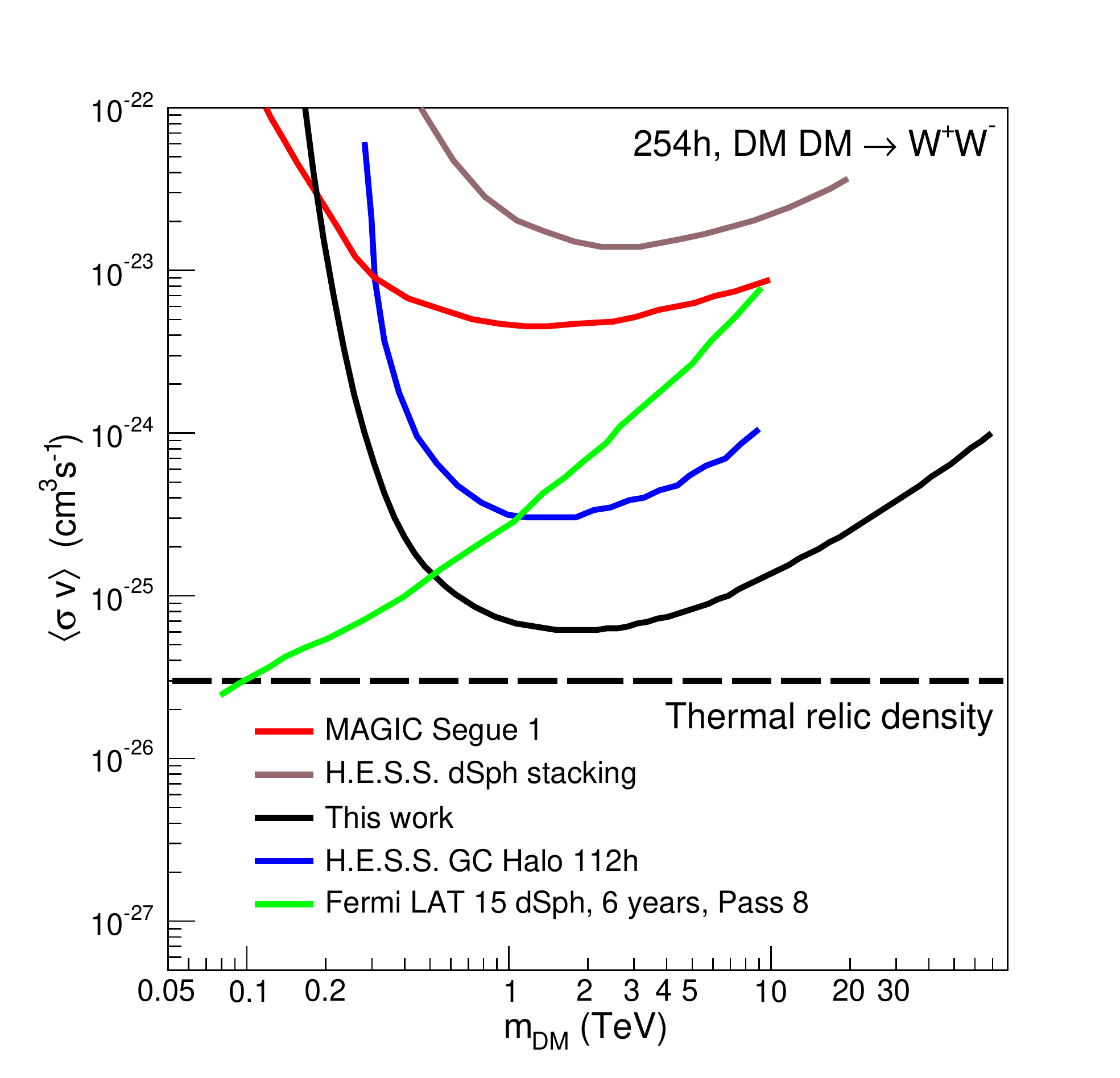}
\caption{Left: Impact of the DM density distribution on the constraints on the velocity-weighted annihilation cross section $\langle \sigma v \rangle$. The constraints expressed in terms of 95\% C. L. upper limits are  shown 
as a function of the DM mass m$_{\rm DM}$ in the W$^+$W$^-$ channels
for the Einasto profile (solid black line), another parametrization of the Einasto profile (dotted black line), and the NFW profile (long dashed-dotted black line), respectively.  
Right: Comparison of constraints on the W$^+$W$^-$ channels with the previous published H.E.S.S. limits from 112 hours of observations of the GC~\cite{Abramowski:2011hc} (blue line), the limits from the observations of 15 dwarf galaxy satellites of the Milky Way by the Fermi satellite~\cite{Ackermann:2015zua} (green line), the limits from 157 hours of observations of the dwarf galaxy Segue 1~\cite{Aleksic:2013xea} (red line), and the combined analysis of observations of 4 dwarf galaxies by H.E.S.S.~\cite{Abramowski:2014tra} (brown line).}
\label{fig:SummaryPlot}
\end{figure*}
We find no significant $\gamma$-ray excess in any of the ON regions (RoIs) with respect to the OFF regions~\cite{supplement}. We derive upper limits on $\langle \sigma v \rangle$  at 95\% C. L. for WIMPs with masses from
160 GeV to 70 TeV, annihilating into quark ($b\bar{b}$, $t\bar{t}$), gauge boson (W$^+$W$^-$) and lepton ($\mu^+\mu^-$, $\tau^+\tau^-$) channels.
The $\gamma$-ray spectrum from DM annihilation in the channel $f$ is computed by using the tools available from Ref.~\cite{Cirelli:2010xx}. The left panel of Fig.~\ref{fig:results_errorJ} shows the observed 95\% C. L. upper limits for the W$^+$W$^-$ channel and the Einasto profile. The expected limits are obtained from 1000 Poisson realizations of the background obtained through observations of blank fields at high latitudes where no signal is expected (see Supplemental Material~\cite{supplement}).
The mean expected upper limit together with the 68\% and 95\% containment bands are plotted. The limits reach $6 \times 10^{-26}$ cm$^3$s$^{-1}$ for a DM particle of mass 1.5 TeV. We obtain a factor of five improvement compared with the results of Ref.~\cite{Abramowski:2011hc}. 
The larger dataset and the improved data analysis method contribute to the increase of the sensitivity of the analysis presented here. In the right panel of 
Fig.~\ref{fig:results_errorJ}, the observed 95\% C. L. upper limit is shown for the  $\tau^+\tau^-$ channel
together with the expected limits. The limits reach $\langle \sigma v \rangle$ values expected for 
dark matter annihilating at the thermal-relic cross section. 
The observed upper limits together with expectations are given for the $b\bar{b}$, $t\bar{t}$ and $\mu^+\mu^-$ channels, respectively, in Fig.~\ref{fig:results_channels} in Supplemental Material~\cite{supplement}. 
The limits obtained in the leptonic  channels ($\mu^+\mu^-$, $\tau^+\tau^-$) are comparatively strong with respect to those in the quark channels ($b\bar{b}$, $t\bar{t}$). This mainly comes from the relatively soft measured $\gamma$-ray spectra compared to the hard ones stemming from leptonic annihilation channels. 
In the left panel of Fig.~\ref{fig:SummaryPlot}, the impact of the DM distribution hypothesis on the observed upper limit is shown for the NFW profile and an alternative parametrization of the Einasto profile extracted from Ref.~\cite{Cirelli:2010xx}.

The right panel of Fig.~\ref{fig:SummaryPlot}  shows a comparison with the current constraints obtained from the observations of the MAGIC ground-based Cherenkov telescope instrument towards the Segue 1 dwarf galaxy~\cite{Aleksic:2013xea}\footnote{The J-factor of Segue 1 used in Ref.~\cite{Aleksic:2013xea} could be overestimated by a factor of 100 as shown in Ref.~\cite{Bonnivard:2015xpq}.},  the combined analysis of 4 dwarf galaxies observed by H.E.S.S.~\cite{Abramowski:2014tra}, and  the observations of 15 dwarf galaxy satellites of the Milky Way by the Fermi satellite~\cite{Ackermann:2015zua}.

\section{Summary}
We present a new analysis of the inner halo of the Milky Way using 10 years of observation of the GC (254 h of live time) by the phase 1 of H.E.S.S.  and a novel statistical analysis technique using a 2D maximum likelihood method. We find no evidence of a gamma-ray excess and thus exclude
a velocity-weighted annihilation cross section of $\rm 6 \times 10^{-26}$ cm$^3$s$^{-1}$ for DM particles with a mass of  1.5 TeV annihilating in the $W^+W^-$ channel for an Einasto profile. 
These are the most constraining limits obtained so far in the TeV mass range. 
Our constraints surpass the Fermi limits for particle masses above 400 GeV in the $W^+W^-$ channel. The strongest limits are obtained in the $\tau^+\tau^-$ channel at $\rm 2\times10^{-26}\ cm^3s^{-1}$ for a DM particle mass of 1 TeV. For the first time, observations with a ground-based array of imaging atmospheric Cherenkov telescopes are able to probe the thermal relic annihilation cross section in the TeV DM mass range.

The upcoming searches with H.E.S.S. towards the inner Galactic halo will benefit from additional observations of the phase 2 of H.E.S.S. which aims for an energy threshold lowered down to several tens of GeV and improved sensitivity in the TeV energy range. Higher Galactic latitude observations will allow increasing the source region size and in turn reduce the impact of the uncertainty of the DM distribution in the inner kpc of the GC. 
Within the next few years, searches with H.E.S.S.  observations are expected to explore in-depth 
the WIMP paradigm for TeV DM particles.

\section{Acknowledgments}
The support of the Namibian authorities and of the University of Namibia in facilitating the construction and operation of H.E.S.S. is gratefully acknowledged, as is the support by the German Ministry for Education and Research (BMBF), the Max Planck Society, the German Research Foundation (DFG), the French Ministry for Research, the CNRS-IN2P3 and the Astroparticle Interdisciplinary Programme of the CNRS, the U.K. Science and Technology Facilities Council (STFC), the IPNP of the Charles University, the Czech Science Foundation, the Polish Ministry of Science and Higher Education, the South African Department of Science and Technology and National Research Foundation, the University of Namibia, the Innsbruck University, the Austrian Science Fund (FWF), and the Austrian Federal Ministry for Science, Research and Economy, and by the University of Adelaide and the Australian Research Council. We appreciate the excellent work of the technical support staff in Berlin, Durham, Hamburg, Heidelberg, Palaiseau, Paris, Saclay, and in Namibia in the construction and operation of the equipment. This work benefited from services provided by the H.E.S.S. Virtual Organisation, supported by the national resource providers of the EGI Federation.

\bibliography{bibl}
\clearpage
\appendix
\setcounter{equation}{0}
\setcounter{figure}{0}
\widetext
\begin{center}
{\bf \large \large Supplemental Material: Search for a Dark Matter annihilation signal\\ towards the inner Galactic halo with H.E.S.S.}
\end{center}
\section{Background measurement technique}
\begin{figure*}[!h]
\centering
\includegraphics[width=0.6\textwidth]{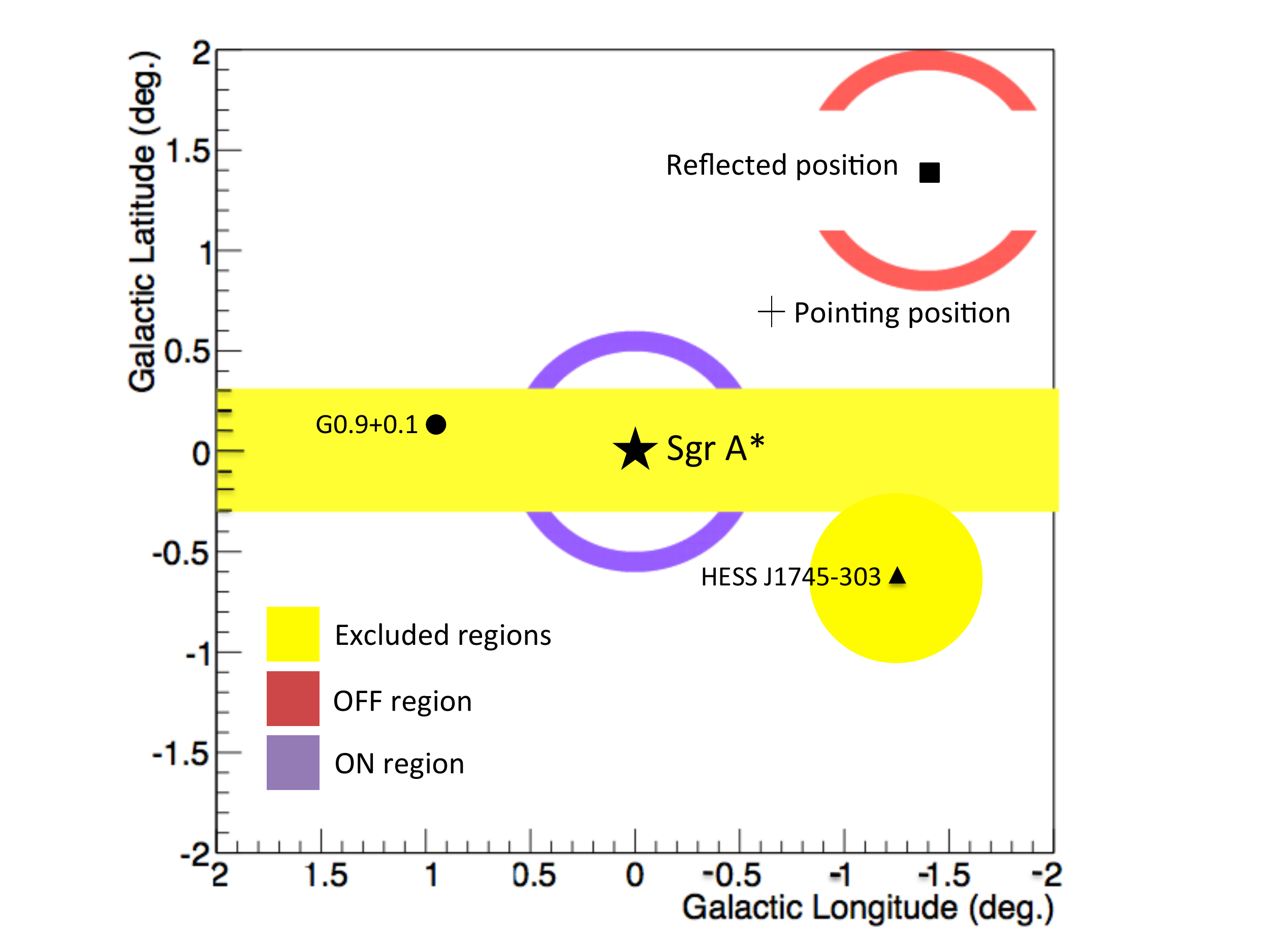}
\caption{Schematic of the background measurement technique for a pointing position at (-0.7$^{\circ}$,+0.7$^{\circ}$) in Galactic coordinates. The OFF region (red-filled open ring) is taken symmetrically to the ON region (blue-filled open ring) from the observational pointing position (black cross). 
By construction, ON and OFF regions have the same angular size. The positions of Sgr A* (black star), G0.9+0.1 (black dot) and HESS J1745-303 (black triangle) are shown. The yellow-filled box with Galactic latitudes from -0.3$^{\circ}$ to +0.3$^{\circ}$ and the yellow-filled disc are excluded for signal and background measurements. }
\label{fig:background_method}
\end{figure*}

The background is measured for each pointing position 
 of the GC observations with the H.E.S.S. instrument in the same camera field of view as for the signal. The GC dataset is composed of about 600 observational runs with pointing positions taken between Galactic longitudes and latitudes of -1.1$^{\circ}$ and +1.1$^{\circ}$, respectively. For a given pointing position of the H.E.S.S. array, the background is measured in an OFF region taken symmetrically to the ON  region from the pointing position as in Ref.~\cite{Abramowski:2011hc}. This enables a determination of the expected background in the ON region from a measurement in the OFF region taken under the same observational and instrumental conditions as for the signal measurement in the ON region. 
All regions of the sky with $\gamma$-ray sources (yellow-filled regions in Fig.~\ref{fig:background_method}) are excluded for ON and OFF measurements. 
By construction, the ON and OFF regions have the same angular size. Fig.~\ref{fig:background_method} shows a schematic of the background measurement technique for a pointing position at (-0.7$^{\circ}$,+0.7$^{\circ}$) in Galactic coordinates and for the region of interest of inner and outer radii of 0.5$^{\circ}$ and 0.6$^{\circ}$, respectively. For a pointing position inside the ON region, the region which intersects the ON and the OFF regions is excluded. 

\section{ON and OFF event distribution in the regions of interest}
The analysis of ground-based Cherenkov telescope data utilizes the ON-OFF method to search for a $\gamma$-ray excess above the background in the signal region. As mentioned above, the background events in the OFF region are measured at the same time as the events in the ON region within the camera field of view. This background measurement technique on a run-by-run basis provides an accurate determination of the $\gamma$-ray  background in the ON region. The signal region, defined as a circle of 1$^{\circ}$ radius centered in the GC with Galactic latitudes between $\pm$0.3$^{\circ}$ excluded, is divided into seven open annuli of 0.1$^{\circ}$ width.  
No $\gamma$-ray excess is found between the ON and OFF regions in any of the RoIs. The numbers of ON and OFF events summed over all the energy bins, are compatible within 1$\sigma$ for any of the seven RoIs. The absence of excess between the measured number of $\gamma$-ray events in ON and OFF regions allow us to derive constraints on $\langle \sigma v \rangle$.

\section{Statistical methodology}
\begin{figure*}[!h]
\centering
\includegraphics[width=0.5\textwidth]{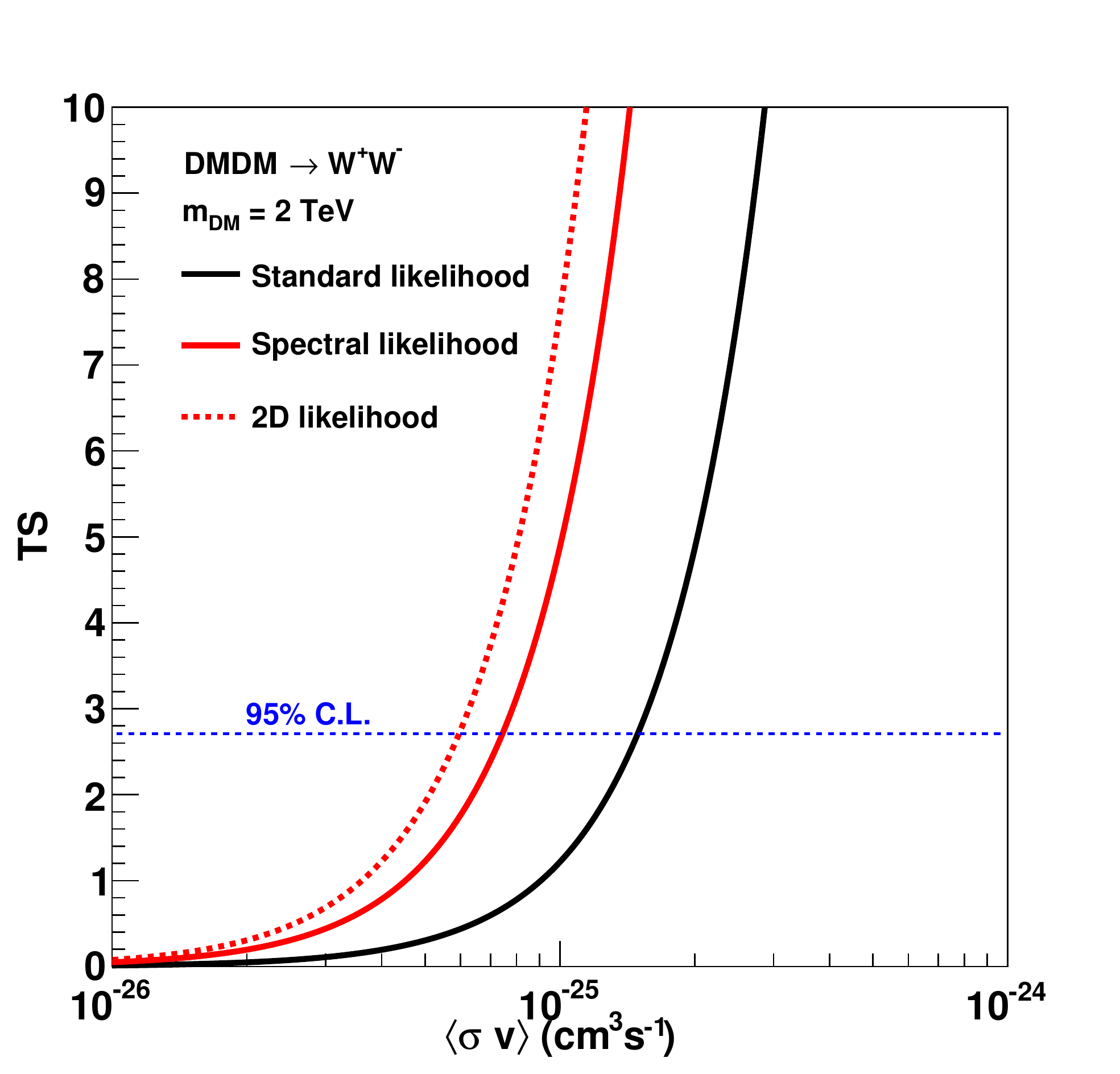}
\caption{TS profile for the standard (solid black line), spectral (solid red line) and  2D (dotted red line) likelihoods, respectively,  for DM particles of 2 TeV mass annihilating into W$^+$W$^-$ pairs. The 95\% confidence level upper limit requires a change in the test statistic value of 2.71 (dotted blue line). The standard likelihood approach corresponds to the method employed in the previous analysis of the inner Galactic halo~\cite{Abramowski:2011hc}.
}
\label{fig:likelihood}
\end{figure*}

The statistical analysis is based on a maximum likelihood formalism to test the DM signal hypothesis and derive constraints on $\langle  \sigma v \rangle$ for a set of DM mass $m_{\rm DM}$ and annihilation spectrum $\rm d N_{\gamma}/\rm d E_{\gamma}$. The constraints on $\langle  \sigma v \rangle$ are computed via a test statistics  TS requiring a change of 2.71 for the TS profile to obtain 95\% C.L. upper limits.
The formalism makes use of the spatial and spectral characteristics of the DM signal and the 2D likelihood function is given by the product of Poisson likelihoods over the spatial and energy bins. 
In the 2D likelihood, the number of events can be summed 
over the spatial bins and over both the spatial and spectral bins to obtain the {\it spectral likelihood} and {\it standard likelihood} formalisms, respectively. The standard likelihood approach has been employed in the 
previous analysis of the inner Galactic halo~\cite{Abramowski:2011hc}.
The 2D likelihood approach provides improved sensitivity over the latter two. Fig.~\ref{fig:likelihood} shows the improvement of 2D likelihood with respect to the standard and spectral likelihoods  for a DM particle of mass of 2 TeV annihilating in W$^+$W$^-$ pairs. The spectral approach profits of the peculiar spectral feature of the DM signal compared to background and an increase in the sensitivity of a  factor of $\sim$1.9 is obtained. In addition to the spectral behavior, the spatial dependence of the DM signal with respect to background  
can be used. The increase in sensitivity  with the 2D likelihood is $\sim$30\% compared to the spectral likelihood.

\section{J-factor}
\label{sec:jfactor}
The regions of interest consist of seven annuli of 0.1$^{\circ}$ width and centered at the GC, with inner radii from  0.3$^{\circ}$ to 0.9$^{\circ}$. A band 
from Galactic latitudes $b = -0.3^{\circ}$ to $b = +0.3^{\circ}$ is excluded along the Galactic plane.  
Table~\ref{tab:jfactors} 
provides the inner and outer radii of the RoIs together with their angular size and J-factor values for Einasto and NFW profiles. 
\begin{table}[h]
\centering
\begin{tabular}{c|c|c|c|c|c|c}
$i^{\rm th}$ RoI & Inner radius & Outer radius & Angular size &\multicolumn{3}{c}{$J$-factor}  \\
 &  [deg.] &  [deg.] & [10$^{-5}$ sr] &\multicolumn{3}{c}{[10$^{20}$ GeV$^2$cm$^{-5}$]}  \\
\hline
&  &&& Einasto &  NFW & Einasto~\cite{Cirelli:2010xx} \\
\hline
\hline
1& 0.3 & 0.4 & 4.9& 4.3 & 2.5 & 1.3\\
2& 0.4 & 0.5 &6.8 & 5.6 & 3.0 & 1.7\\
3 & 0.5 & 0.6 &8.7 & 6.6 & 3.3 & 2.0 \\
4 & 0.6 & 0.7 &10.6 & 7.4 & 3.5 &  2.3 \\
5& 0.7 & 0.8 &12.5 & 7.9 & 3.6 &  2.5 \\
6 & 0.8 & 0.9 & 14.4& 8.3 & 3.7 &  2.6 \\
7 & 0.9& 1.0 & 16.3& 8.7 & 3.8 &  2.8 \\
\hline
\hline
\end{tabular}
\caption{J-factor values in units of GeV$^2$cm$^{-5}$ in each of  the 7 RoIs considered in this work. The first four columns give the inner radius, the outer radius, and the size in solid angle for each RoI. The fifth column provides the J-factor values  for the Einasto profile considered in this work together with the values obtained for a NFW profile~\cite{Abramowski:2011hc} and an alternative normalization of the Einasto profile~\cite{Cirelli:2010xx}. \label{tab:jfactors}}
\end{table}

\section{Expected limit computation}
The expected limit in the inner GC region is calculated from H.E.S.S. observations in high-Galactic-latitude ($\rm |b|>10^{\circ}$) fields where all the H.E.S.S. sources are excluded. For each observational data taking run of the GC dataset, the expected background is selected from the blank-field-observation database for the observational conditions of the run (zenith angle, offset, muon efficiency). For a given observational run which lasts about 28 minutes, the residual background extracted from the database in the run observational conditions is available for higher time exposure It is then renormalized to the duration of the considered run. 
It defines the mean expected background. A Poisson realization from the mean expected background is calculated for each observational run and an overall realization of the expected background for the whole GC dataset is obtained when summing the realizations over all the runs. This procedure is repeated 1000 times over all the observational runs. For each DM mass, the mean, hereby referred to as expected, the 68\% and 95\% containment bands of the distribution of the $\langle \sigma v \rangle$ values are computed. 

\section{Flux sensitivity}
Using Eq.~(1), the predicted $\gamma$-ray flux integrated above 100 GeV
can be computed as function of the DM mass, and compared to the expected and observed flux sensitivities from 
\begin{figure}[!h]
\centering
\includegraphics[width=0.5\textwidth]{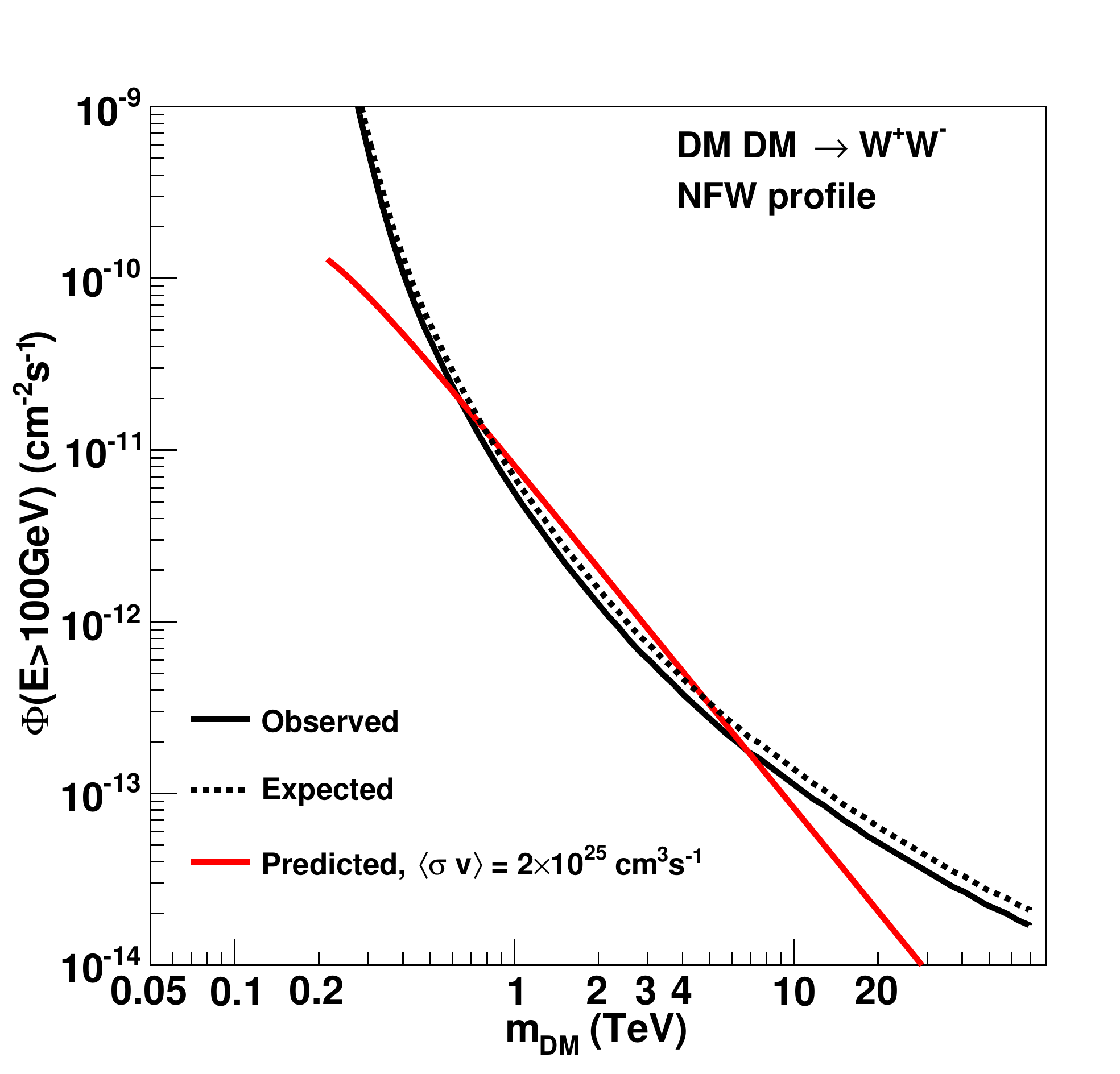}
\caption{Expected (solid black line) and observed (dashed black line) $\gamma$-ray flux sensitivities above 100 GeV at 95\% C. L. as a function of the DM mass for DM particles annihilating into $W^+W^-$ pairs and the NFW profile. The predicted flux above 100 GeV for $\langle  \sigma v \rangle$ = 2$\times$10$^{-25}$ cm$^3$s$^{-1}$
is also shown (solid red line).}
\label{fig:flux}
\end{figure}
10 years of GC observations by H.E.S.S.  The expected and observed upper limits at 95\% C.L. on $\langle  \sigma v \rangle$ can be used to derive expected and observed sensitivities  at 95\% C.L on the $\gamma$-ray flux.
Fig.~\ref{fig:flux} shows the predicted flux above 100 GeV for DM particles annihilating into $W^+W^-$ pairs with 
$\langle  \sigma v \rangle$ = 2$\times$10$^{-25}$ cm$^3$s$^{-1}$ and the NFW profile. The H.E.S.S. expected and observed flux sensitivities at 95\% C. L. for the  $W^+W^-$ channel are shown. For the NFW profile and the $W^+W^-$ channel, a $\langle  \sigma v \rangle$ value of 2$\times$10$^{-25}$ cm$^3$s$^{-1}$ can be probed for DM masses between 800 GeV and 6 TeV given the H.E.S.S. flux sensitivity in the inner 1$^{\circ}$ on the Galactic center.

\section{Annihilation channels}
\label{sec:channels}
\begin{figure*}[!h]
\centering
\includegraphics[width=0.32\textwidth]{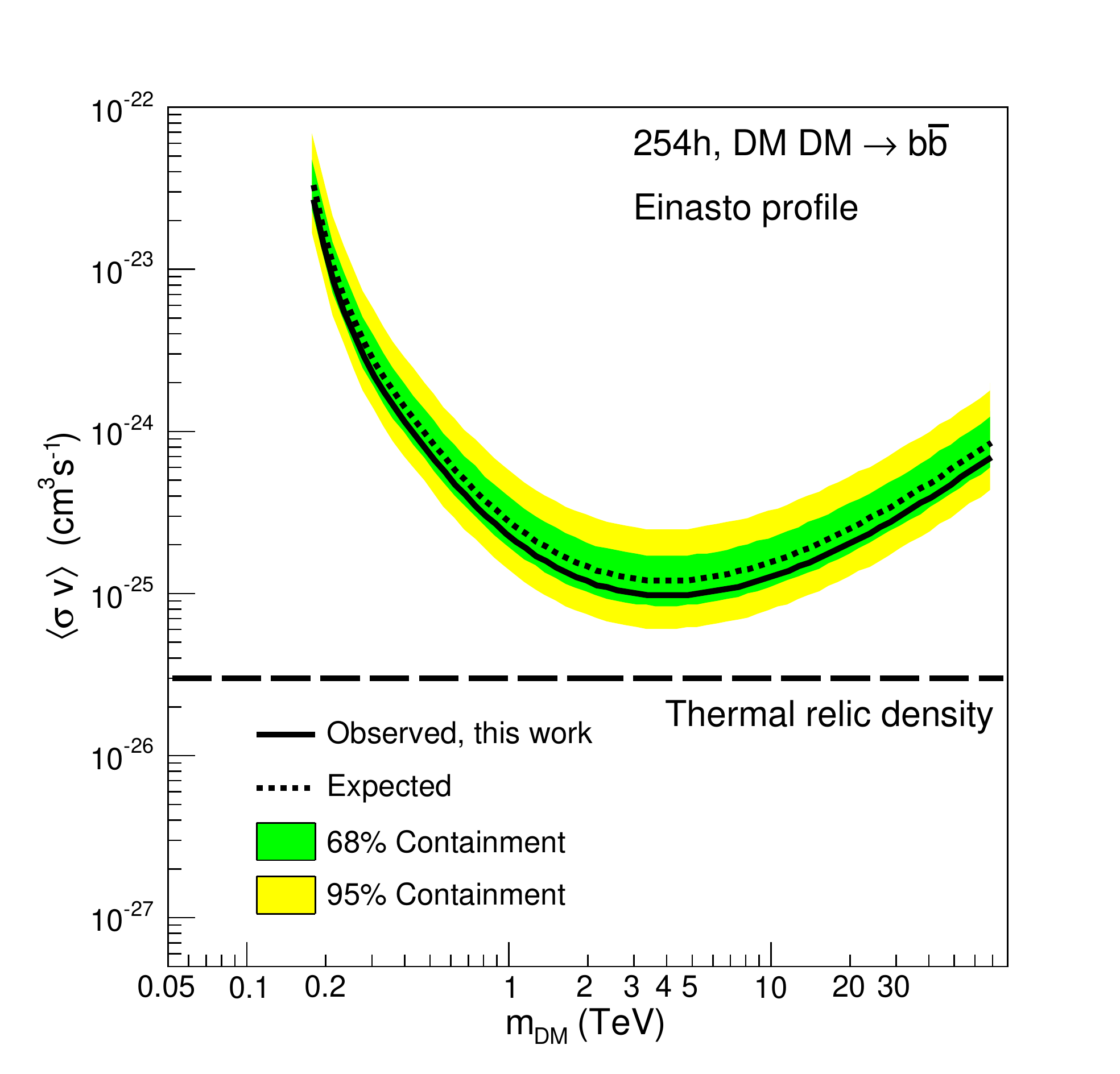},
\includegraphics[width=0.32\textwidth]{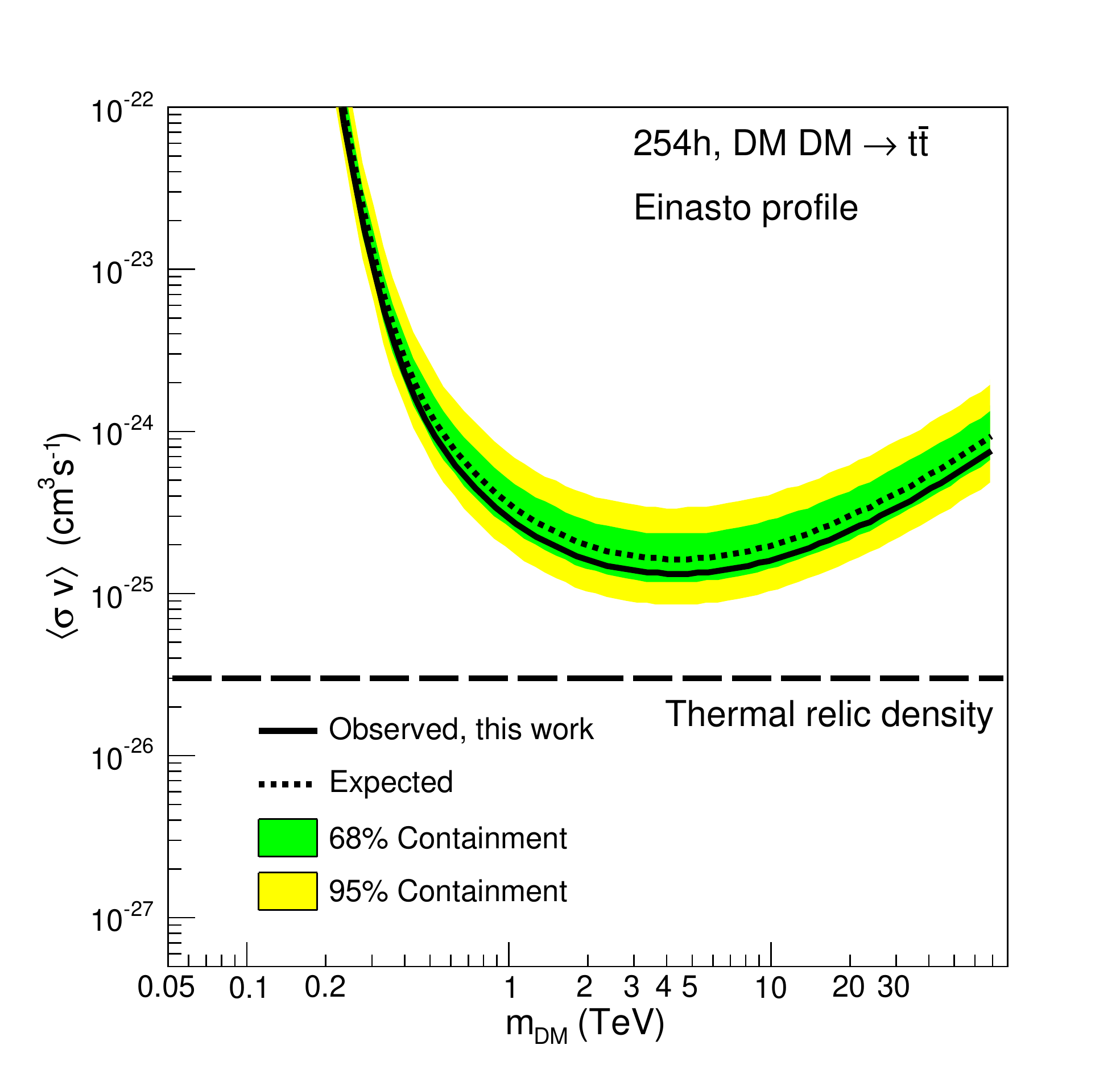},
\includegraphics[width=0.32\textwidth]{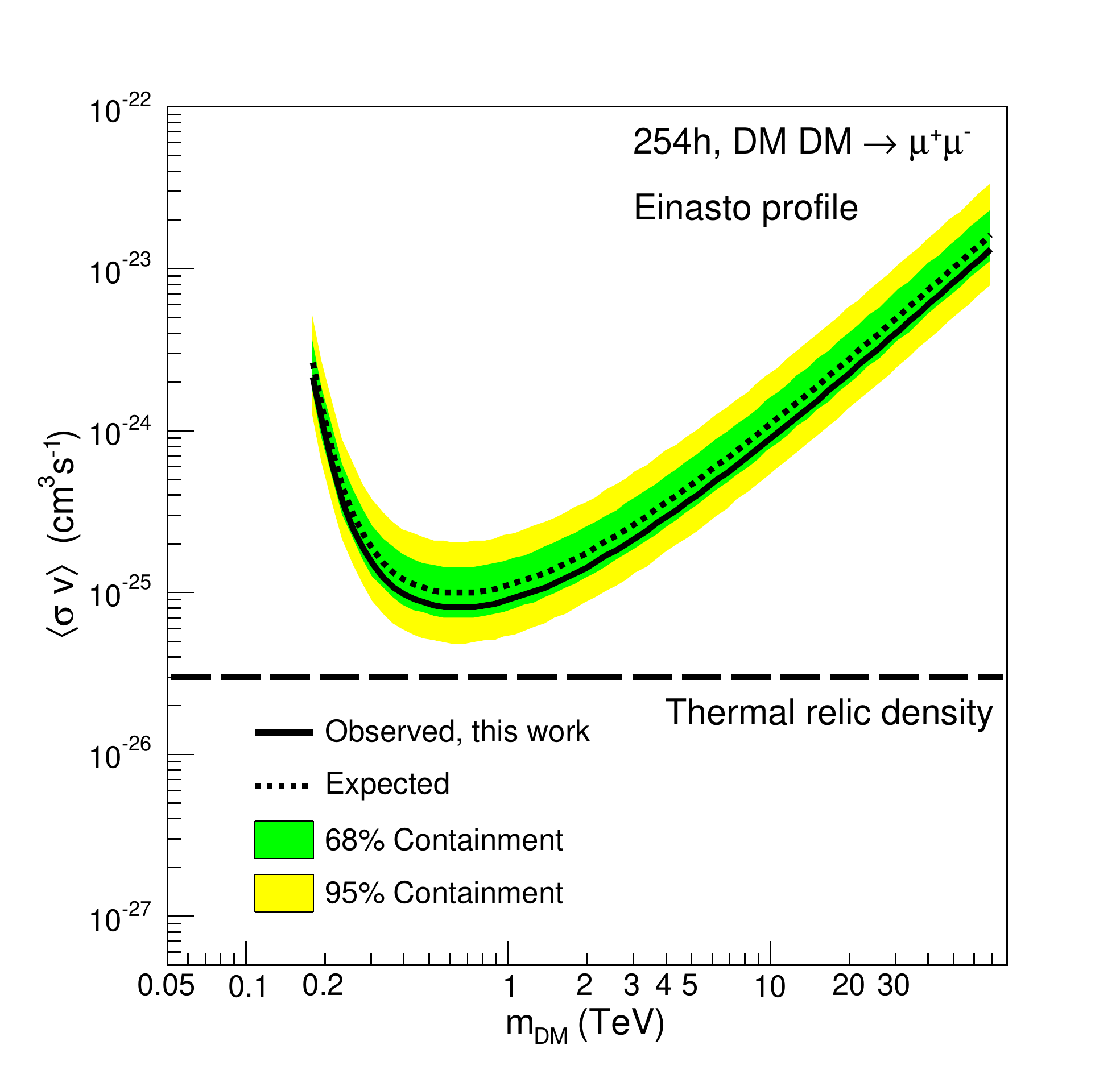}
\caption{Constraints on the velocity-weighted annihilation cross section $\langle \sigma v \rangle$ 
derived from 10 years of GC observations by H.E.S.S. for the  $b\bar{b}$ (left panel),  $t\bar{t}$ (central panel) and $\mu^-\mu^+$ (right panel) channel, respectively. The constraints are expressed in terms of 95\% C. L. upper limits as a function of the DM particle mass $m_{\rm DM}$. 
The observed upper limit is plotted as a solid black line. Expected limits are derived from blank-field observations at high Galactic latitudes. 
The mean expected limit (black dashed line) is plotted together with the 65\% (green) and 95\% (yellow) confidence bands.
The natural scale for $\langle \sigma v \rangle$ is given as long-dashed black line.}
\label{fig:results_channels}
\end{figure*}

WIMPs can self-annihilate into pairs of standard model particles allowed by kinematics. For the DM mass range considered in this work, the light-quark ($u\bar{u}$, $d\bar{d}$, $c\bar{c}$, $s\bar{s}$, $b\bar{b}$) and the three lepton (e$^+$e$^-$, $\mu^+\mu^-$, $\tau^+\tau^-$) channels are always open. The $t\bar{t}$ channel is kinematically open as soon as $m_{\rm DM} \ge m_{\rm t}$. We performed the analysis in the $\mu^+\mu^-$, $\tau^+\tau^-$, $b\bar{b}$, $t\bar{t}$ and $W^+W^-$ channels, in each case assuming 100\% branching ratio. The constraints are shown in Fig.~\ref{fig:results_channels} for the $b\bar{b}$, $t\bar{t}$, and $\mu^+\mu^-$, respectively.
\end{document}